\documentclass[intlimits,twoside,a4paper]{article}

\usepackage[cp1251]{inputenc}
\usepackage{xcolor}
\usepackage[eqsecnum]{cmpj3}

\usepackage{bm}

\newcommand{\braket}[2]{\langle \, {#1} \, | \, {#2} \, \rangle}

\def\ket#1{\, | \, {#1} \, \rangle}

\issue{2020}{23}{4}{43710}
\doinumber{10.5488/CMP.23.43710}
\title[Anomalous thermodynamics of a quantum spin system with large residual entropy]%
{Anomalous thermodynamics of a quantum spin system with large residual entropy}%
\author[J. Richter \it{et al.}]{J. Richter\refaddr{label1,label2}, J.
Schulenburg\refaddr{label3}, D.V.~Dmitriev\refaddr{label4},
V.Ya.~Krivnov\refaddr{label4}, J. Schnack\refaddr{label5}} 
\addresses{
\addr{label1} Institut f\"{u}r Physik, Otto-von-Guericke-Universit\"{a}t
Magdeburg, P.O. Box 4120, 39016 Magdeburg, Germany
\addr{label2} Max-Planck-Institut f\"{u}r Physik komplexer Systeme,
N\"{o}thnitzer Stra\ss e 38, 01187 Dresden, Germany
\addr{label3} Universit\"atsrechenzentrum, Universit\"at Magdeburg, D-39016
Magdeburg, Germany
\addr{label4} Institute of Biochemical Physics of RAS,
           4 Kosygin St., 119334 Moscow, Russia
\addr{label5} Fakult\"{a}t f\"{u}r Physik, Universit\"{a}t Bielefeld,
          Postfach 100131, 33501 Bielefeld, Germany
}
\date{Received June 24, 2020, in final form August 14, 2020}

\begin{document}

\maketitle
\begin{abstract}
In contrast to strongly frustrated classical systems, their quantum counterparts typically have a
non-degenerate
ground state. A counterexample is the celebrated Heisenberg sawtooth spin chain with
ferromagnetic zigzag bonds $J_1$ and competing antiferromagnetic basal
bonds $J_2$.
At a quantum phase transition point  $|J_2/J_1|=1/2$, this model exhibits a
flat one-magnon excitation band
leading to a massively degenerate ground-state manifold which
results in a large residual entropy. Thus, for the spin-half model, the
residual
entropy  amounts to exactly one half of its
maximum value $\lim_{T\to\infty} S(T)/N = \ln2$.
In the present paper we study in detail the 
role of the spin quantum number $s$ and  the magnetic field $H$ in the
parameter region
around the transition (flat-band) point.
For that we use full exact diagonalization up to $N=20$ lattice sites and
the finite-temperature Lanczos method up to $N=36$ sites to calculate the
density
of states as well as the temperature dependence of the specific
heat, the entropy and the susceptibility.
The study of chain lengths up to $N=36$  allows a careful finite-size
analysis.
At the flat-band point we find extremely small finite-size effects for spin
$s=1/2$, i.e., the
numerical data virtually correspond to the thermodynamic limit.
In all other cases the finite-size effects are still small and become
visible at
very low temperatures. In a sizeable parameter region around the flat-band
point
 the former
 massively degenerate ground-state manifold acts as 
a large manifold of low-lying excitations leading  to extraordinary
 thermodynamic properties at the transition point
as well as in its vicinity such as an additional
low-temperature maximum in the specific heat.
Moreover, there is a very strong influence of the  magnetic field on the
low-temperature thermodynamics including an enhanced magnetocaloric effect.

\end{abstract}

\section{Introduction}
\label{sec1}
\setcounter{equation}{0}

\begin{figure}[!t]
\begin{center}
\includegraphics[clip=on,width=70mm,angle=0]{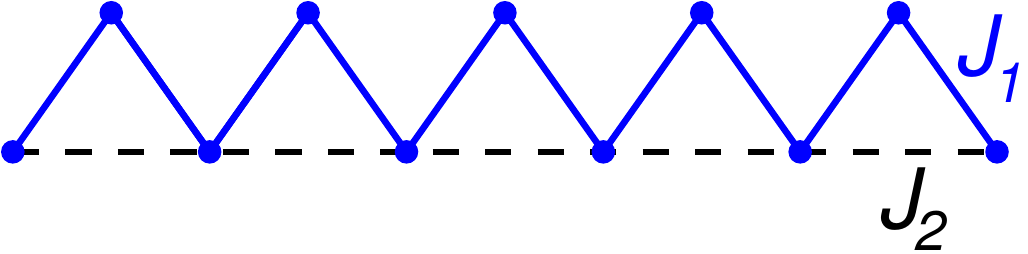}
\caption{(Colour online) A schematic picture of the sawtooth chain.}
\label{F1} 
\end{center}
\end{figure}

The sawtooth chain is one of the paradigmatic frustrated
quantum spin models built
of corner-sharing  triangles.
The corresponding Heisenberg Hamiltonian is given by
\begin{equation}
\label{model}
{\cal H}
=
J_1\sum_{\langle i,j\rangle } {\bf s}_{i} \cdot  {\bf s}_{j}
+ J_2 \sum_{\langle\langle i,j\rangle\rangle }{\bf s}_{i} \cdot {\bf s}_{j} 
- H\sum_{i}S_i^z
\end{equation}
with ${\bf s}_i^2=s(s+1)$. Here, the first sum runs over the zigzag bonds and
the second one runs over the basal bonds, see figure~\ref{F1}.     
There are numerous studies of this spin model, see, e.g.,
references~\cite{1,2,3,4,5,6,7,8,9,10,11,12,13,14,15,16,17,18,19,20,21,22,23,24}
within different contexts  ranging from exact dimer product ground states
\cite{1,3,4,14} via
quantum three-coloring  description
\cite{17,18,23}
to many-body quantum
scars \cite{24}.
As a  prototype of a flat-band model,
the sawtooth chain has attracted a particular attention by  the  community
investigating frustrated quantum spin systems, see, e.g.,
references~\cite{2,9,10,12,13,15,16,19,21,23}
as well as by  groups studying electronic
systems, see, e.g.,  
references~\cite{25,26,27,28,29,30,31,32,33},
and also photonic lattices, see, e.g.
references~\cite{34,35}.
Further motivation for theoretical studies comes from several magnetic
compounds where the magnetic ions reside on sawtooth lattice sites, see,
e.g.  
references~\cite{36,37,38,39,40}.

The focus of the present paper is on a specific version of the sawtooth
Heisenberg model with 
ferromagnetic (FM) zigzag bonds $J_1<0$ and competing antiferromagnetic (AFM) basal bonds
$J_2>0$. We call  this model the FM-AFM sawtooth chain. 
This model undergoes  a quantum phase transition at $\kappa_\text{c}= |J_2/J_1|=1/2$
 from a FM to a ferrimagnetic ground state
 \cite{7,11,15},
where $\kappa_\text{c}$ is
 independent of the spin quantum number $s$ \cite{15}.
At the transition (flat-band) point $\kappa_\text{c}$,  
the lowest one-magnon excitation band from the FM state becomes
flat and it has zero energy
\cite{15,16,19,22,23}.
Such a flat one-magnon band leads to a massively degenerate set  of localized multi-magnon
ground states resulting in an
$s$-independent residual entropy $\lim_{N \to \infty}  S_0(N)/N = \frac{1}{2}\ln 2$
\cite{15}, which is even larger than for the 
AFM sawtooth chain at  its flat-band point \cite{9,10}.
While the thermodynamics of the AFM sawtooth chain  is well studied
see, e.g., references~\cite{9,10,12,13,41},
so far only a few investigations are available for the
 FM-AFM sawtooth chain
\cite{15,16,19,22,23}.
Since the flat-band physics for this model can be observed at zero field, it
might be  even more interesting than the AFM model, where 
flat-band physics appears around the saturation field, i.e., typically at high
magnetic fields.
Therefore, in contrast to the  FM-AFM sawtooth chain,
the high-field flat-band physics for purely AFM models is not easily (or not at all) accessible  
in experiments, for the few exceptions see references~\cite{42,43,44,45,46}.
A strong further motivation to extend the theoretical study of the
FM-AFM sawtooth chain comes from the recently synthesized magnetic molecule
Gd$_{10}$Fe$_{10}$
\cite{38}.
This magnetic system is well described by  the 
FM-AFM sawtooth chain with
of $10+10$ alternating gadolinium
($S=\frac{7}{2}$) and iron ($S=\frac{5}{2}$) ions, sitting on apical and
basal sites, correspondingly. Importantly, the ratio  of its exchange parameters
is close to the transition point 
\cite{16,19,38}.
We further mention that the model
is also relevant
for Cs$_2$LiTi$_3$F$_{12}$  that hosts ferro-antiferromagnetic sawtooth chains as magnetic subsystems
\cite{47}.

Here, we present a systematic study of the role of the system size $N$, the
spin quantum number $s$ as well as the change of the thermodynamic properties in
dependence  on the
distance to the transition point $d_\text{f}=|J_2/J_1|-\kappa_\text{c}$.
Moreover, we also discuss the influence of the magnetic field on the
thermodynamics, which may have a strong impact on the low-temperature physics,
because it partially lifts  the huge degeneracy present at $
|J_2/J_1|=\kappa_\text{c}$.  
To this end, we use full exact diagonalization (ED) and the finite-temperature Lanczos
method (FTLM).

\section{Methods}
\label{methods}
\subsection{Full exact diagonalization (ED)}
The exact-diagonalization
technique  is a powerful numerical tool which is
widely applied 
to  quantum lattice models, see, e.g. reference~\cite{48}. 
Using a complete set of basis states, the stationary Schr\"odinger
equation for a finite system of $N$ sites 
is transformed into 
an eigenvalue problem. Then, the full spectrum can be determined 
by numerical diagonalization  without approximations. We use here J\"org
Schulenburg's {\it spinpack}
code
\cite{49,50} which allows one to easily treat  
periodic $s=1/2$ sawtooth chains up to $N=20$ sites.

\subsection{Finite-temperature Lanczos method (FTLM)}
\label{FTLM}

The FTLM is a Monte-Carlo like extension of the full ED
briefly described in the previous section.
Thermodynamic quantities are determined using trace estimators
\cite{51,52,53,54,55,56,57,58,59,60,61, 61a}.
The partition function $Z$ is approximated by a Monte-Carlo like representation of $Z$, 
i.e., the sum over a complete set of $(2s+1)^N$ basis states entering $Z$
is replaced by a much smaller sum over $R$ random vectors $\ket{\nu}$ for each subspace ${\mathcal H}(\gamma)$ of the Hilbert
space.
To split the Hilbert
space into small subspaces 
we use conservation of total $S^z$  as well as  the lattice symmetries of the
Hamiltonian, where  
the mutually orthogonal subspaces are labeled by $\gamma$. 
The exponential of the Hamiltonian is approximated by its spectral representation in a Krylov space 
spanned by the $N_\text{L}$ Lanczos vectors starting from the respective random vector $\ket{\nu}$. 
Then, the FTLM approximation of the  partition function is given by 
\begin{eqnarray}
\label{315}
Z(T)
\approx
\sum_{\gamma=1}^\Gamma\!
\frac{\text{dim}({\mathcal H}(\gamma))}{R}
\sum_{\nu=1}^R\!
\sum_{n=1}^{N_\text{L}}\!
\exp\left(\!-\frac{\epsilon_n^{(\nu)}}{kT}\!\right)\!\vert\braket{n(\nu)}{\nu}\vert^2
.
\end{eqnarray}
Here, $\ket{n(\nu)}$ is the $n$-th eigenvector of ${\cal H}$ in the Krylov space with the corresponding energy $\epsilon_n^{(\nu)}$.
As for ED, we use  
Schulenburg's {\it spinpack} code
\cite{49,50} for numerical Lanczos calculations to get FTLM data for periodic
sawtooth chains up to $N=36$ sites.

\section{Results}
\subsection{The model at the flat-band point}
Let us first recapitulate some important results found in
reference~\cite{15}. Due to the characterization of the ground-state
manifold by localized multi-magnon (LMM) states, explicit expressions were found for
the degeneracies in each $S_z$-sector.    
For periodic spin-half chains with
$N$ sites, the ground-state degeneracy at the transition point in a particular $S_z$-sector
with
$S^{z}=S_{\max }-k$   
is 
\begin{eqnarray}
D_{N}^{k} &=&\frac{n!}{(n-k)!(k)!}\,,\qquad 0\leqslant k\leqslant \frac{n}{2} \, , \;\ n=\frac{N}{2}\,, \nonumber \\
D_{N}^{k} &=&\frac{n!}{(n/2)!(n/2)!}+\delta _{k,n} \, , \qquad \frac{n}{2}<k\leqslant n .
\label{number}
\end{eqnarray}
This yields 
the total degeneracy 
\begin{equation}
D_N=2^{n}+n\frac{n!}{(n/2)!(n/2)!}+1, \;\; n=\frac{N}{2}
\end{equation}
leading to 
 a residual entropy per site 
$s_0 = \lim_{N\to\infty} \frac{1}{N}\ln D_N = \frac{1}{2}\ln 2$. Note that this value
is independent of $s$ (for $N\to\infty$ only) and it
 corresponds to a system of $N/2$ independent spin-half objects.  
Interestingly, the excitation gap $\Delta$ above the ground-state manifold is extremely
small. Thus, for 
$N=20$
the ED yields
$\Delta = 7.502 \cdot 10^{-9}$.
Moreover, $\Delta $ decreases with increasing $N$.
Thus, the FM-AFM sawtooth chain at $|J_2/J_1|=\kappa_\text{c}$ is a (rare) example of a virtually gapless finite quantum spin system.
 
Applying a magnetic field $H>0$,  the fully polarized state with
$S_z=S_{\max }=Ns$ becomes the ground state and the former LMM ground states of
the other $S_z$-sectors are
excited states, where their excitation energy is related to the Zeeman term. 
Due to  their huge degeneracy, see equation~(\ref{number}),
this class of excitations may dominate the low-temperature physics.
Thus, the contribution of the LMM states to the partition function can be
explicitely given, see equation~(27) in reference~\cite{15}.   
Based on this  knowledge, 
universal scaling relations for the magnetization and the susceptibility
were found.
For the susceptibility, the universal finite-size
scaling function
reads \cite{15}
\begin{equation}
\chi_{N}(T)=T^{-\alpha }f(c_NNT^{\alpha -1}),
 \label{chiscal}
\end{equation}
where the $N$-dependent factor $c_N$ (given by a cumbersome 
formula) becomes  $c_N=1/48 $ for $N\gg1$ and 
the scaling exponent $\alpha$ for $s=1/2$ was determined to $\alpha \simeq 1.09$ by
fitting to corresponding finite-size data for $N=16$ and $N=20$.
Below we  verify this scaling behaviour by comparing the   
 corresponding finite-size data for much larger systems up to $N=36$.

\begin{table}[!b]
\begin{centering}
\caption{Excitation gaps $\Delta(k)\equiv
E_1(k)-E_0$
of the periodic FM-AFM sawtooth chain 
of $N=16$ sites at the transition point
$|J_2/J_1|=1/2=\kappa_\text{c}$, $J_1=-1$, for spin quantum numbers $s=1/2$, $1$, $3/2$ and $2$
in different subspaces $S^z=Ns-k$.
$E_1(k)$ is the energy of the lowest excitation in the subspace of $k$ magnons and
$E_0=-12s^2$ is the ground-state
energy. 
\label{tab1}}
\vspace{2ex}
\begin{tabular}{|c||c|c|c|c|}
\hline
& $s=1/2$ & $s=1$  & $s=3/2$  & $s=2$ \tabularnewline 
\hline 
 $k$  & $\Delta(k)$ &   $\Delta(k)$ & $\Delta(k)$ & $\Delta(k)$   \tabularnewline
 \hline 
 1            & $1.0                $  &  $ 2.0                $   &  $ 3.0                $  & $ 4.0                $    \tabularnewline
 2            & $2.178\cdot 10^{-2}$  &  $ 7.094\cdot 10^{-2}$   &  $ 8.596\cdot 10^{-2}$  & $  9.326\cdot 10^{-2}$    \tabularnewline
 3            & $4.718\cdot 10^{-4}$  &  $ 4.829\cdot 10^{-3}$   &  $ 6.902\cdot 10^{-3}$  & $  7.924\cdot 10^{-3}$    \tabularnewline
 4            & $9.935\cdot 10^{-6}$  &  $ 2.740\cdot 10^{-4}$   &  $ 4.797\cdot 10^{-4}$  & $  5.917\cdot 10^{-4}$    \tabularnewline
 5            & $3.034\cdot 10^{-6}$  &  $ 1.550\cdot 10^{-4}$   &  $ 2.682\cdot 10^{-4}$  & $  3.248\cdot 10^{-4}$    \tabularnewline
 6            & $2.584\cdot 10^{-6}$  &  $ 1.550\cdot 10^{-4}$   &  $ 2.682\cdot 10^{-4}$  & $  3.248\cdot 10^{-4}$    \tabularnewline
 7            & $7.361\cdot 10^{-7}$  &  $ 1.550\cdot 10^{-4}$   &  $ 2.682\cdot 10^{-4}$  & $  3.248\cdot 10^{-4}$    \tabularnewline
\hline                                                         
\end{tabular}                                                  
\par\end{centering}
\end{table}

\subsubsection{Density of states}
\label{DOS}
Before we  present our numerical data for thermodynamic quantities such as
the specific heat $c(T)$ and the susceptibility $\chi(T)$,
we briefly illustrate the low-energy spectrum
by discussing the excitation gaps and the density of states $\rho(E)$.
In table~\ref{tab1}
we present ED data for the excitation gaps in different sectors  of $S^z=Ns-k$
for $N=16$ and  spin $s=1/2$, $1$, $3/2$ and $2$.
Obviously, the gaps are rather small if
$k>1$, where the extreme quantum case plays a particular role with a
virtually vanishing gap (see also   table~I in reference~\cite{15} 
for $s=1/2$ with $N=16,20,24,28$ and $k=1,\ldots,6$).

\begin{figure}[!b]
	\vspace{-.5cm}
	\centering
	\includegraphics[width=.98\textwidth]{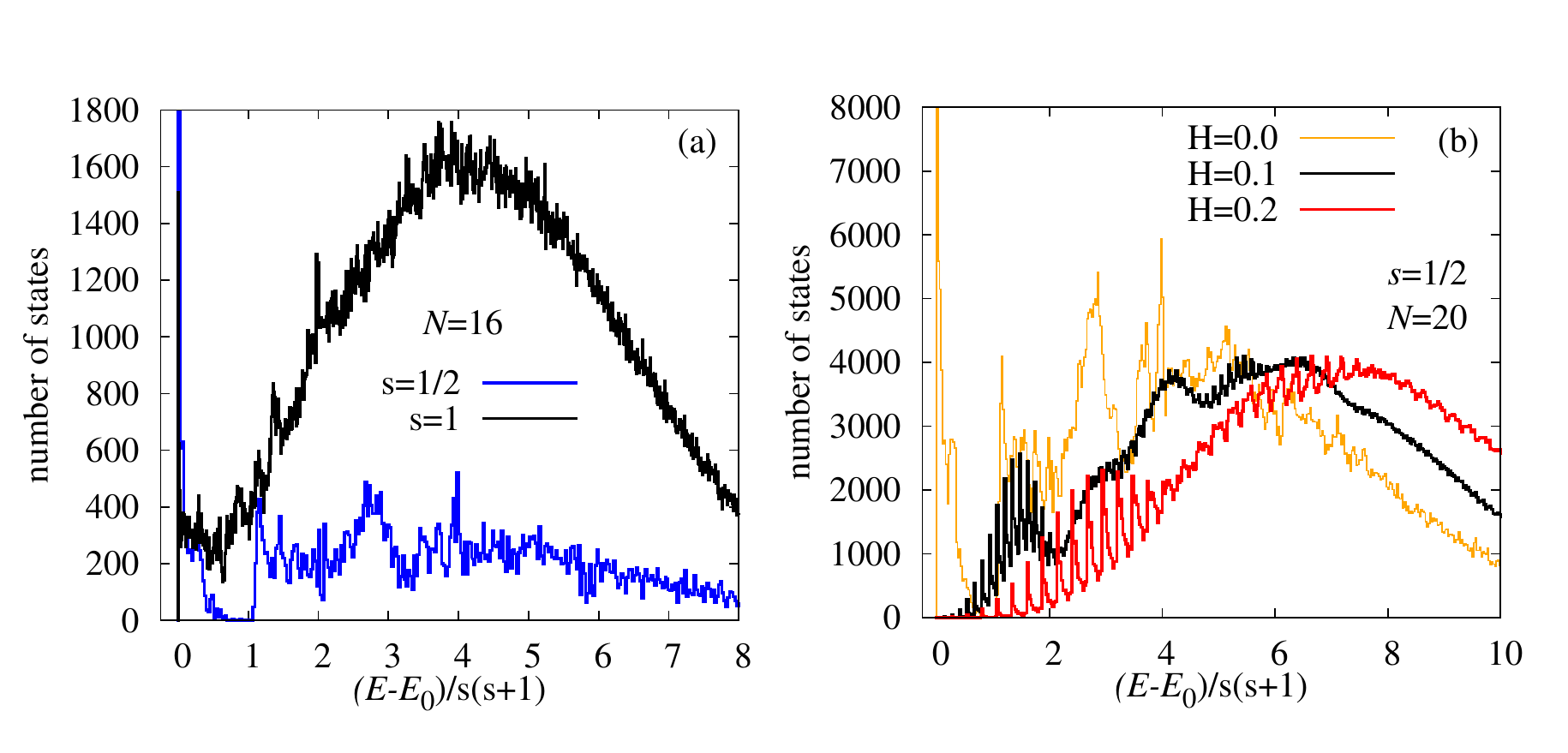}
	\caption{(Colour online) (a) Density of states (histogram, bar width $\Delta E=0.02$) 
		of periodic chains of $N=16$ sites with $J_1=-1$ and $J_2=1/2$ for spin $s=1/2$ and $s=1$
		(ED), where the $y$-axis is cut at $1800$.   
		Note that within the first histogram bar between $E_0$ and $E_0+\Delta E$ 
		not only the ground states but also excited states are collected, see also
		table~\ref{tab1}. 
		(b) Field dependence of the density of states (histogram, bar width $\Delta E=0.02$) 
		of periodic chains of $N=20$ sites with $J_1=-1$, $J_2=1/2$ and spin
		$s=1/2$
		(ED), where the $y$-axis is cut at $8000$.
	}
	\label{F2}
\end{figure}

In figure~\ref{F2}~(a) we show $\rho(E)$ for $N=16$ and spin quantum numbers
$s=1/2$ and  
$s=1$.
An exceptional feature of the  density of states for $s=1/2$ is the collection
of about $6\%$ of the states in the low-energy region below $E-E_0 \lesssim
0.6$, where this region is separated by a quasi-gap from the  high-energy
region  $E-E_0 \gtrsim 0.6$. This feature is also present  for larger system
sizes, see references~\cite{15} and \cite{23}.
The particular low-energy structure  of $\rho(E)$    
is important for the 
low-temperature physics, see below.
As can be also seen in figure~\ref{F2}~(a), the separation of the  low-energy
part of the spectrum is much less pronounced for $s=1$ and gradually vanishes
at further increasing of $s$.
Another peculiar feature of the spectrum of the $s=1/2$ model is the absence of
the expected typical maximum of the density of states in the middle of the
spectrum.

The influence of a small magnetic field on the density of states is
illustrated  in figure~\ref{F2}~(b) for $s=1/2$ and $N=20$.
As already  briefly discussed above, at $H > 0$, the ground state is the single
fully polarized ferromagnetic state and the
degeneracy of the different $S_z$ sectors is lifted.
However, the degeneracy of the LMM states  within a certain $S_z$ sector, see
equation~(\ref{number}),
is still present leading to the unconventional spiked structure of $\rho(E)$ 
below $(E-E_0)/s(s+1) \lesssim 2$ for $H=0.1$    
and
below $(E-E_0)/s(s+1) \lesssim 4$ for $H=0.2$, where the location of the peaks
corresponds to
the Zeeman energy of the respective $S_z$ sector.
These parts of the spectrum related to the LMM states  will dominate the
low-temperature properties.

\begin{figure}[!b]
\vspace{-.5cm}
\begin{center}
\includegraphics[width=.98\textwidth]{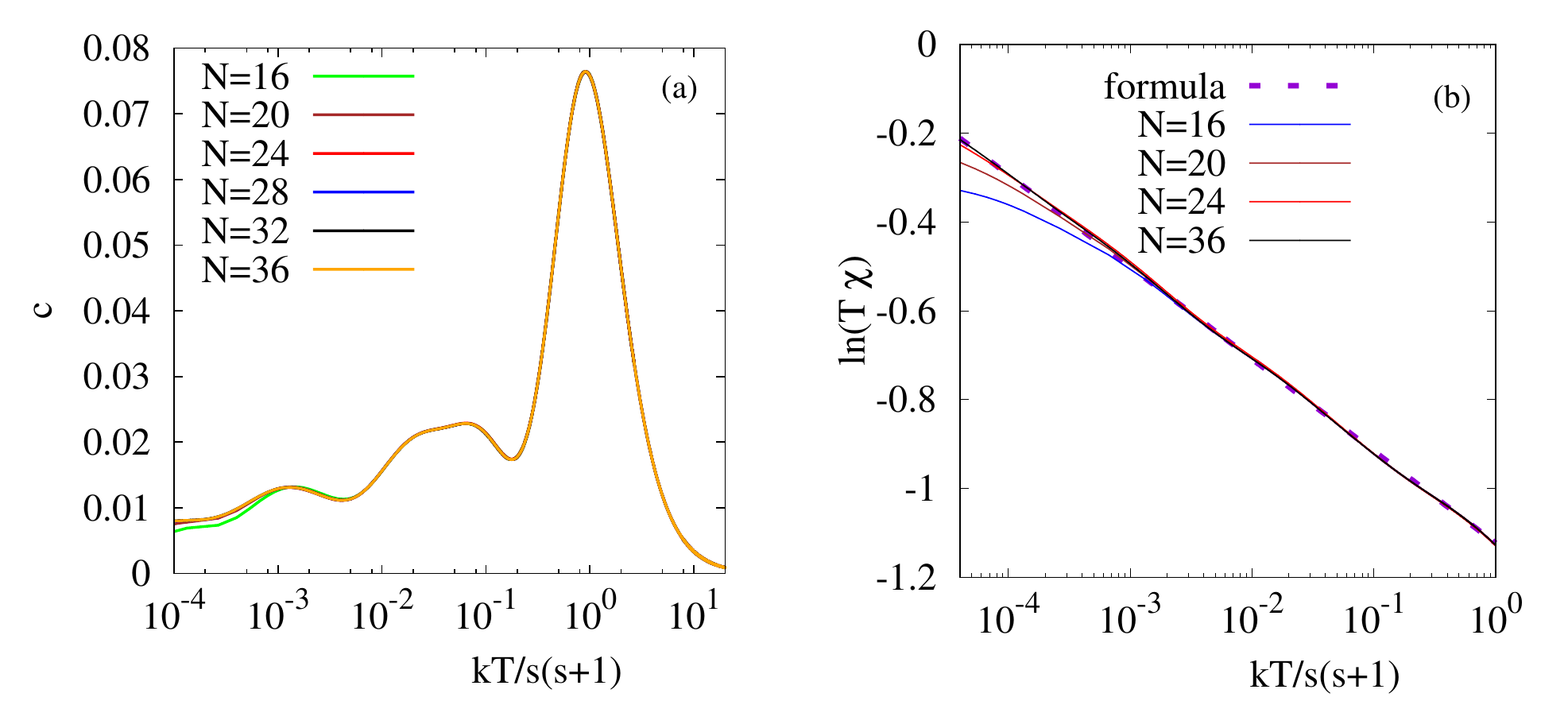}
\caption{(Colour online) (a) Specific heat $c(T)$ per site 
of periodic chains of $N=16,20,24,28,32,36$ sites with $J_1=-1$, $J_2=1/2$ and spin
$s=1/2$ (ED for  $N=16,20$, FTLM for  $N=24,28,32,36$) at zero magnetic
field.
(b) Log-log plot for the dependence of the susceptibility per
site on temperature for periodic chains of $N = 16,20$ (ED) and $N = 24,36$
(FTLM) with $J_1=-1$, $J_2=1/2$ and spin
$s=1/2$. The symbols 
correspond to the formula $\chi(T)=0.317/T^{1.09}$.
}
\label{F3}
\end{center}
\end{figure}

\subsubsection{Specific heat, entropy and uniform susceptibility}
\label{c_and_S}
Similar to the energy scale of the density of states,  for the thermodynamic quantities we use the
normalized temperature $T/s(s+1)$ to get a better comparison between systems of
different $s$. (Note that the temperature dependences of the specific heat $c(T)$
as well as for  the susceptibility
$\chi(T)$ for different~$s$  become identical at high temperatures as a
function of $T/s(s+1)$ \cite{62}.) 

In reference~\cite{15}, 
by comparing data for
$N=16,18,20,22$ for the $s=1/2$ sawtooth chain, it was found  
(i) that  
the low-temperature part is very specific with a long tail down to very low
temperatures including two weak additional maxima below the typical main maximum
and 
(ii) that the finite-size effects seem to be very small.
It is also worth mentioning that the unconventional low-temperature part of $c(T)$ 
below the main maximum is entirely
covered by the energy levels below the quasi-gap, cf.
reference~\cite{23}.

We strengthen these statements by including FTLM data up to $N=36$, see
figure~\ref{F3}~(a), where we show the specific heat at the transition point for
spin $s=1/2$.
Obviously,  
there are  no finite-size effects down to
$T/s(s+1) \sim 0.0001$ (only for the smallest system of 
$N=16$, we see small deviations from the curves for larger $N$ at $T<0.001$). 
We also observe that the FTLM approximation is very accurate, cf.
reference~\cite{60}.
Thus, our finite-size data for spin $s=1/2$ virtually correspond to the thermodynamic limit. 
This feature can be attributed to the virtually vanishing excitation  gaps, cf.
section~\ref{DOS}.  
For the temperature dependence of the susceptibility, which is related to
equation~(\ref{chiscal}), in reference~\cite{15}, the 
formula $\chi(T)=0.317/T^{1.09}$ was found.
In figure~\ref{F3}~(b) we show 
$\ln(T\chi)$ vs. $T/s(s+1)$ for $N=16,20,24,36$.
Obviously,
 the
finite-size effects are again small and the curves for  $N=24$ and $36$
perfectly coincide
with the above given formula for $\chi(T)$
in the whole temperature region shown in figure~\ref{F3}~(b).
Thus, the data for larger $N$ 
further confirm equation~(\ref{chiscal}).  
It turns out that the scaling exponent $\alpha$ present  in the scaling
function  equation~(\ref{chiscal}) depends on the
spin quantum number $s$ \cite{22}. It changes from $\alpha=1.09$ for the extreme quantum case
$s=1/2$
to $\alpha=1.23$ for $s=1$ and then it smoothly tends to $\alpha=1.5$ at $s \to
\infty$.

\begin{figure}[!t]
	\begin{center}
\includegraphics[width=.98\textwidth]{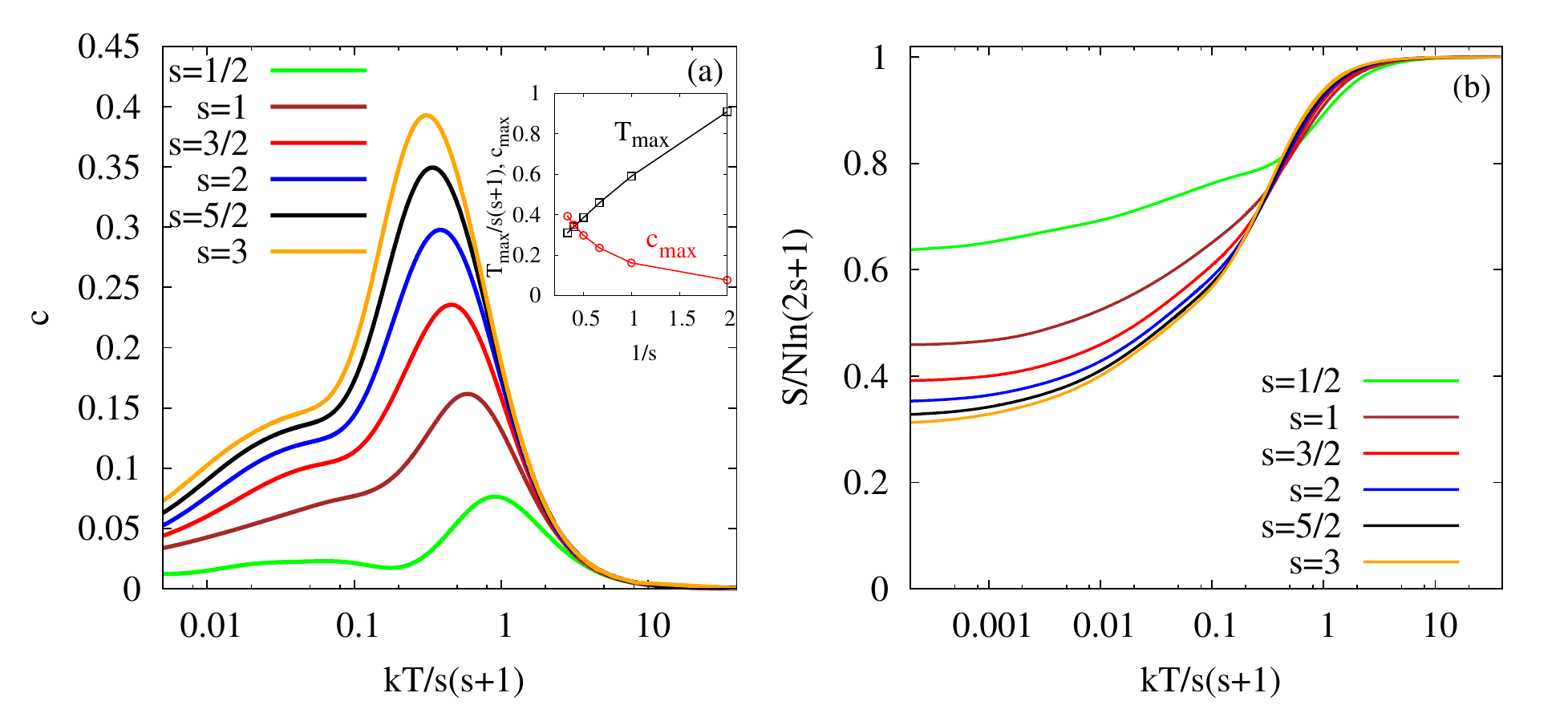}
		\caption{(Colour online) (a) Main panel: Specific heat $c(T)$ per site 
			of periodic chains of $N=12$ sites with $J_1=-1$, $J_2=1/2$ for spin
			$s=1/2,1,3/2,2,5/2,3$ at zero magnetic
			field.
			Inset: Position $T_{\rm max}/s(s+1)$ and height $c_{\rm max}$ of the main maximum
			in $c(T)$ as a
			function of the inverse spin quantum number $s$.  
			(b) Scaled entropy $s(T)/N\ln(2s+1)$ per site 
			for $N=12$ sites with $J_1=-1$, $J_2=1/2$ for spin
			$s=1/2,1,3/2,2,5/2,3$ at zero magnetic
			field. 
		}
		\label{F4}
	\end{center}
\end{figure}

We consider now the sawtooth chain with higher spin $s$.  
In figure~\ref{F4}~(a) and (b) we show the
specific heat and the entropy for $N=12$ 
and spin values $s=1/2, 1, 3/2, 2, 5/2, 3$.
As for $s=1/2$, for all $s>1/2$ we observe a long tail below the main maximum reaching very
low
temperatures.
 However, in contrast to the extreme quantum case $s=1/2$, the   
low-temperature part does not exhibit additional maxima, but rather there is a     
shoulder
just below the main maximum.
The position $T_{\rm max}/s(s+1)$ and the height $c_{\rm max}$ of the main maximum
in $c(T)$ strongly depend on $s$, see the inset of figure~\ref{F4}~(a).
Obviously, the maximum moves to smaller values of  $T/s(s+1)$ at increasing
of $s$.
From the exact solution of the classical  case~\cite{19}
it is known that for $s \to \infty$  there is no maximum,  rather the $c(T)$
exhibits a  plateau-like shape with $c(T<T_\text{p})>0$, $T_\text{p}/s(s+1) \sim 0.2$. While for the most spin systems the low-$T$ thermodynamics for the pretty large
spin value $s=3$
is close to the classical case, we conclude that the highly frustrated FM-AFM sawtooth chain
is an example, where $s=3$  is still far from the classical limit.
For all considered values of $s$, the general  entropy profiles [figure~\ref{F4}~(b)] are similar, though
with different values of the residual entropy.
The slow convergence towards the classical limit with increasing of $s$ may be related
to the  exponentially large ground state degeneracy.

\begin{figure}[!t]
	\begin{center}
\includegraphics[width=.98\textwidth]{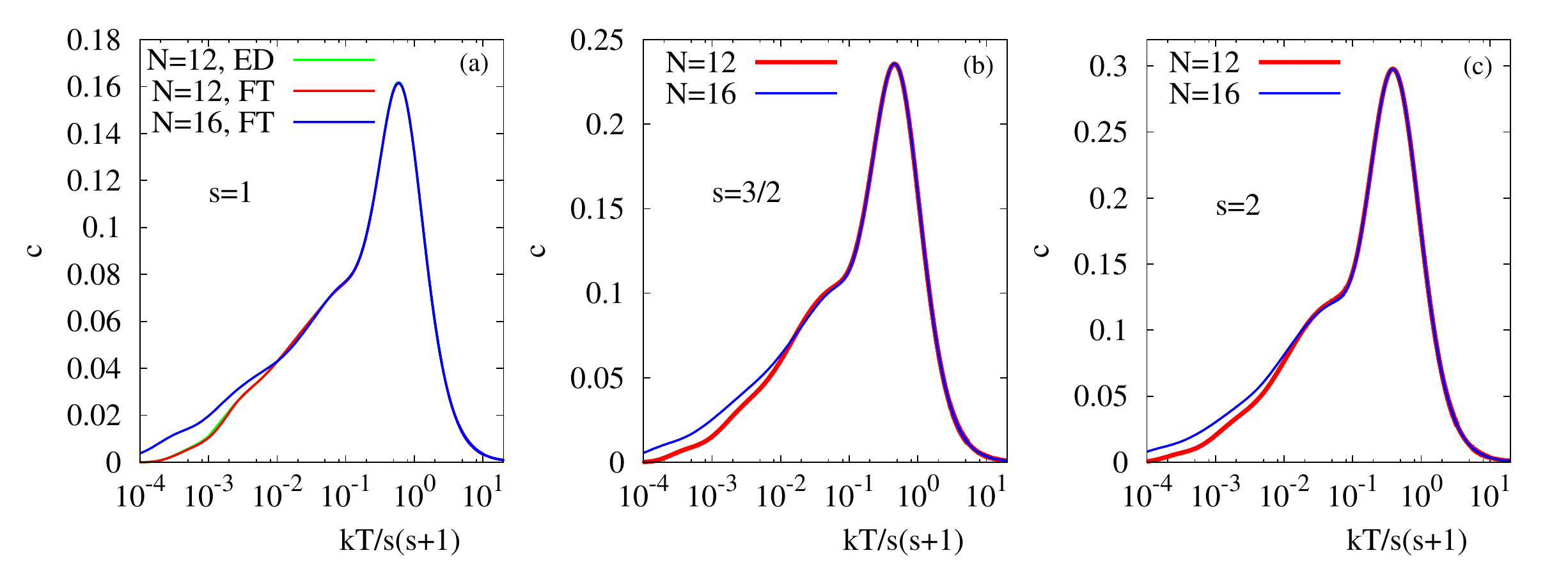}
		\caption{(Colour online) (a) Specific heat $c(T)$ per site 
			for periodic chains of $N=12$ and $16$ sites with $J_1=-1$, $J_2=1/2$ and
			spin
			$s=1$ (ED for  $N=12$, FTLM for  $N=12$ and $N=16$) at zero magnetic
			field. 
			(b)--(c) Same as in (a) but without ED for s = 3/2 and 2,
			respectively. 
		}
		\label{F5}
	\end{center}
\end{figure}

According to the larger excitation gaps for $s > 1/2$, see
table~\ref{tab1},  we may expect that
finite-size effects  set in earlier as $T \to 0$. 
In figure~\ref{F5}~(a), (b), (c) we show the 
specific heat at the flat-band  point $\kappa_\text{c}= |J_2/J_1|=1/2$ for
spin $s=1$, $s=3/2$  and $s=2$  and $N=12$ and $N=16$.
Finite-size effects become visible below the shoulder, i.e., at about  $kT/s(s+1)
< 0.01$, 
but the general shape of the $c(T)$ curve 
 remains similar when increasing $N$.

\begin{figure}[!b]
	\vspace{-.3cm}
	\begin{center}
\includegraphics[width=.98\textwidth]{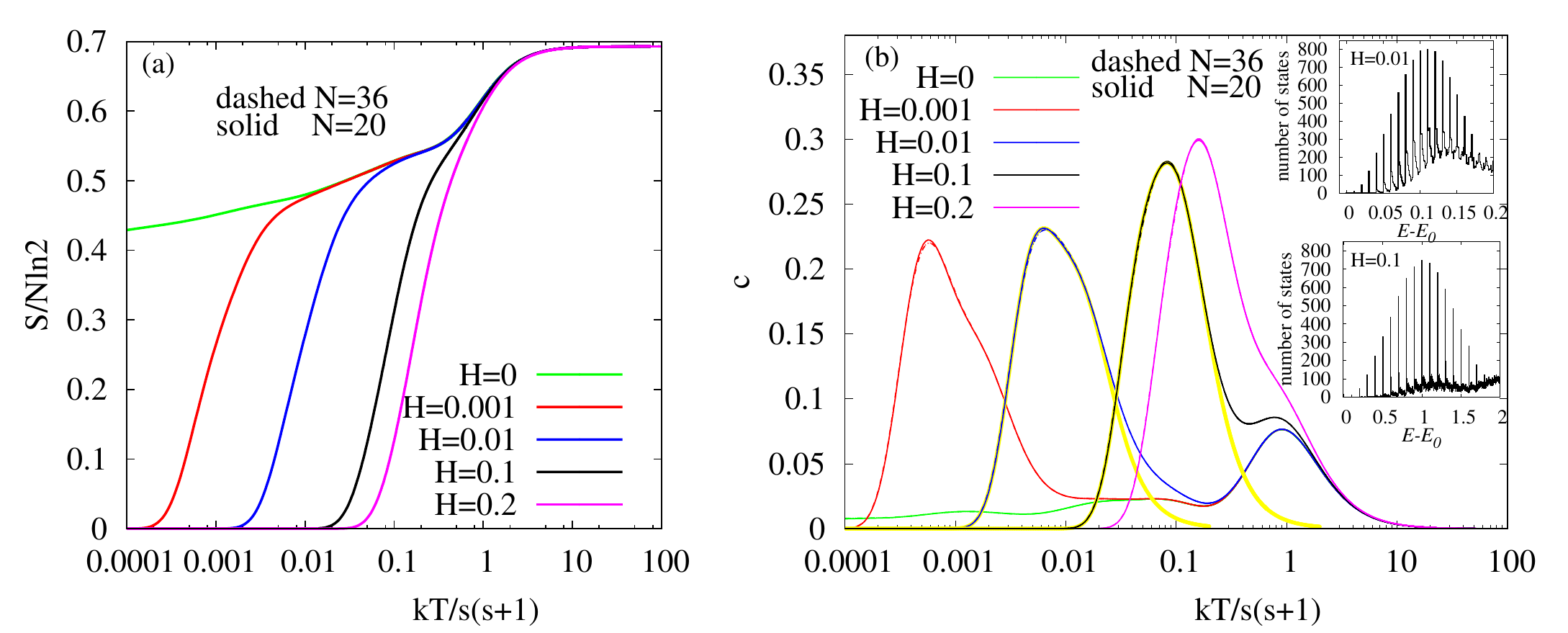}
		\caption{(Colour online) (a) Scaled entropy $s(T)/N\ln(2s+1)$ per site 
			for periodic chains of $N=20$ (ED) and $36$ (FTLM) sites with $J_1=-1$, $J_2=1/2$ and spin
			$s=1/2$ for various magnetic fields $H$.
			(Note that the curves for  $N=20$ and $N=36$ coincide, i.e., the dashed lines
			are practically not visible.)
			(b) Main panel: Influence of the magnetic field $H$ on the specific heat $c(T)$ per site 
			for periodic chains of $N=20$ (ED --- solid) and $N=36$ (FTLM --- dashed) sites with $J_1=-1$, $J_2=1/2$ and spin
			$s=1/2$. (Note that the curves for  $N=20$ and $N=36$ perfectly coincide, i.e., the dashed lines  
			are practically not seen.) The
			broad yellow curves 
			are calculated using a
			restricted set of energies to determine  the specific heat, namely  $E < E_0+0.15$ ($E < E_0+1.5$) for $H=0.01$
			($H=0.1$).
			Inset: Density for states for $H=0.01$ and $H=0.1$ shown for that
			energy region relevant for the extra low-temperature maximum in $c(T)$
			presented
			in the main panel.
		}
		\label{f5}
	\end{center}
\end{figure}

\subsubsection{Influence of the magnetic field}

As briefly discussed for the density of states (see section~\ref{DOS}), a magnetic field may have a
drastic influence on the low-energy physics by partial lifting the massive ground-state
degeneracy. 
Here, we focus on the specific heat $c(T)$ and the entropy $s(T)$.
The magnetization process was recently
studied in detail in reference~\cite{22}.

We present numerical data for  $s(T)$ and $c(T)$ for $N=20$, $N=36$ and
$s=1/2$  in
figure~\ref{f5}~(a) and
(b), respectively, where magnetic fields $H=0,\;0.001,\;0.01,\;0.1,\;0.2$ are
considered. 
The lifting of the massive ground-state degeneracy by the magnetic field is well
visible in figure~\ref{f5}~(a). 
There is no residual entropy at $H>0$, though by increasing the
temperature at small fields, the entropy $s(H,T)$ pretty fast approaches the
zero-field value  $s(H=0,T)$ [green line in figure~\ref{f5}~(a)].

At small nonzero field $H \lesssim 0.121$,  the specific heat exhibits a fairly  high low-temperature maximum. 
This maximum is caused by a series of low-lying excitations stemming
from the former degenerate LMM zero-field ground states, see the well
separated sharp peaks in the density of states shown in the inset of
figure~\ref{f5}~(b). This becomes evident by the broad yellow 
curves which
are determined using a
restricted set of energies to compute the specific heat.
Clearly, the position of the low-temperature maximum $T_{\rm max}(H)$ is related to the
Zeeman
energies of the LMM states and
it is
approximately given by  $T_{\rm max} = 0.615 H$.
Again, finite-size effects are negligible.

\subsubsection{Signs  of flat-band
physics away from the flat-band point $\kappa_\text{c}$}

Realization of the ideal flat-band physics in an experiment on a
sawtooth magnet is unlikely. Rather, one may expect that in a specific magnetic compound,
the exchange parameters are sufficiently close to the flat-band point.
A striking        
example is the FM-AFM sawtooth-chain magnetic molecule Gd$_{10}$Fe$_{10}$ \cite{38}, where
the ratio of exchange
parameters $J_1$ and $J_2$ 
is slightly  below the flat-band  point.
However, the Gd and Fe ions carry large spins $s$ and the system of $N=20$ spins
is not accessible by ED or FTLM.
Hence, we focus here on spin $s=1/2$ that again allows the analysis of
finite-size effects for $|J_2/J_1| \ne \kappa_\text{c}$ by considering various $N$ up
to $N=36$.

The LMM states stemming from the flat-band are
exact eigenstates only at the flat-band point.
Moving away from this point, the states and thus also the eigenvalues are
modified. As a result, the
massive ground-state
degeneracy is lifted and the majority of the former LMM states  become
a large manifold of low-lying excitations. 
One may expect that the change of energies is smooth, where the
excitation energies depend on the distance from the flat-band point
 $d_\text{f}=|J_2/J_1|-\kappa_\text{c}$.  

\begin{figure}[!b]
\begin{center}
\includegraphics[width=.98\textwidth]{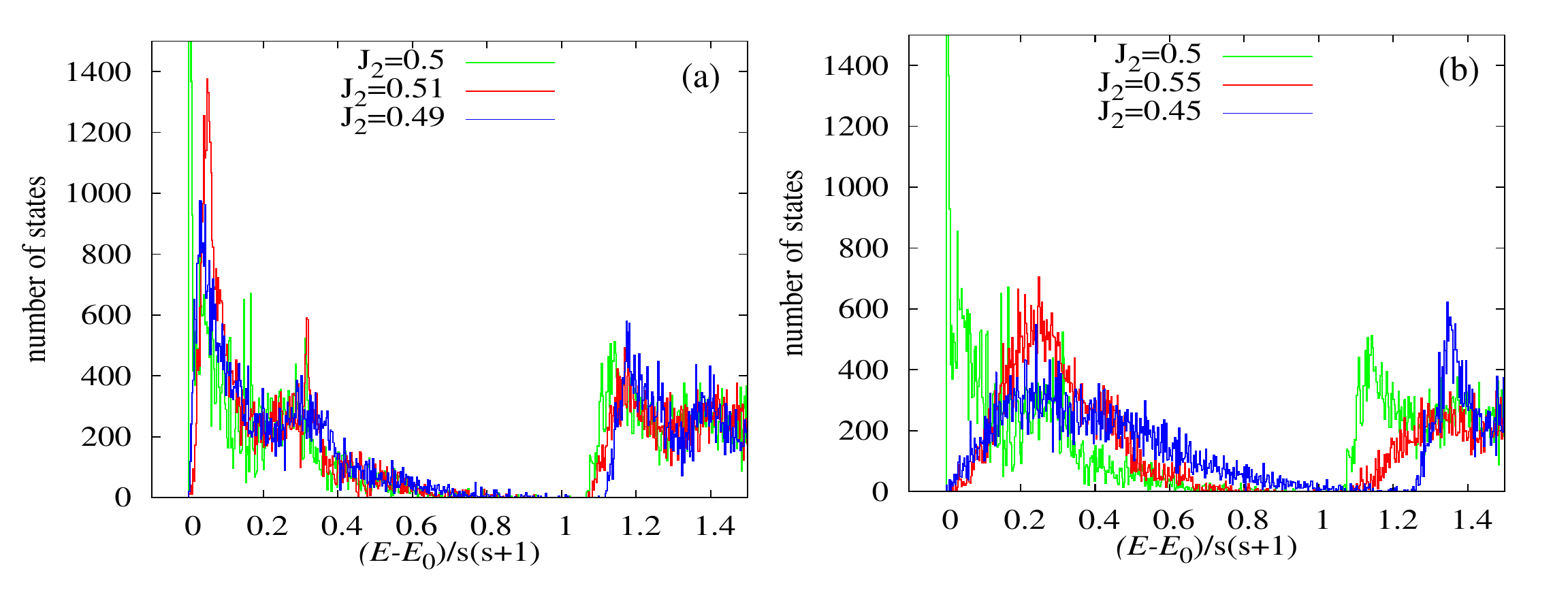}
\caption{(Colour online) Low-energy part of the density of states (histogram, bar width $\Delta
E=0.002$) 
of periodic  $s=1/2$ chains of $N=20$ sites with $J_1=-1$ and 
(a) $J_2=0.49$ and
$J_2=0.51$ as well as 
(b) $J_2=0.45$ and
$J_2=0.55$ compared with  the density of states 
at the flat-band point ($J_2=0.5$). Note that the $y$-axis is cut at $1500$.   
}
\label{f6}
\end{center}
\end{figure}

\begin{figure}[!t]
\begin{center}
\includegraphics[width=.98\textwidth]{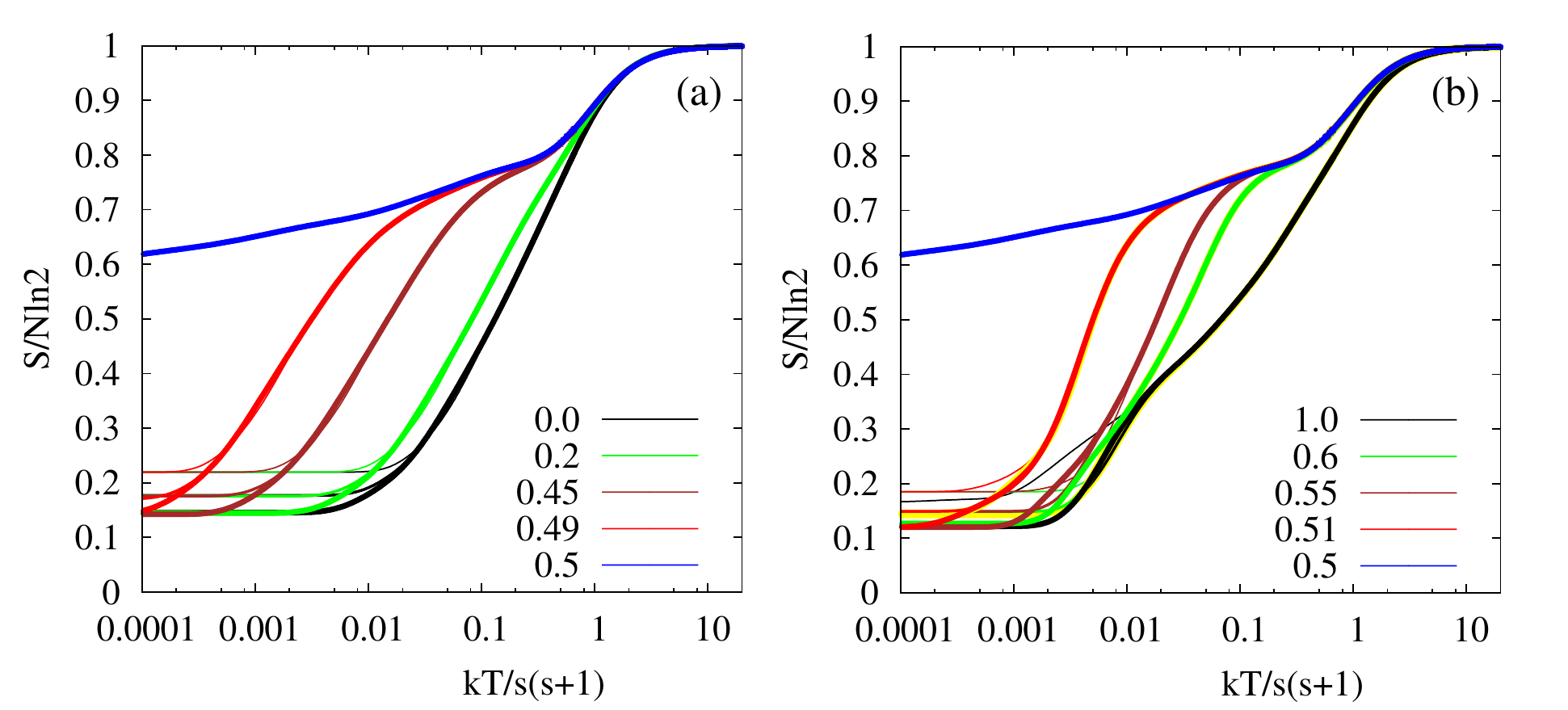}
\caption{(Colour online) Scaled entropy $S(T)/N\ln2$ per site 
of periodic chains of $N=20,28,36$ sites [$N=20$ (ED)~---~thin,
$N=28$ (FTLM) --- medium, $N=36$ (FTLM) --- thick] 
for spin $s=1/2$, $J_1=-1$ and various values
of (a) $J_2
\leqslant \kappa_\text{c}$ and (b) 
$J_2  \geqslant \kappa_\text{c}$ 
(the $J_2$ values are given in the legend)
at zero magnetic field.
}
\label{F8}
\end{center}
\end{figure}

We start with the discussion of the density of states  $\rho(E)$, see
figure~\ref{f6},
where we show the low-energy part of  $\rho(E)$ for several values of $J_2$.
The lifting of the ground-state degeneracy as well as the low-energy manifold of
the former LMM ground states and  their energy shift with growing 
$d_\text{f}=|J_2/J_1|-\kappa_\text{c}$ is evident. Moreover, the quasi-gap is still
present and the states below the quasi-gap determine the low-temperature
physics. 

The specific features of the low-energy spectrum lead to a specific behaviour of
the entropy shown in  figure~\ref{F8}.
There is only a small residual entropy related to the ferromagnetic
(ferrimagnetic) ground-state multiplet at  $J_2<1/2$ ($J_2>1/2$), which
vanishes as $\ln N/N$ when $N \to \infty$. This size-dependent residual
entropy yields the splitting of the curves for various $N$ at low
$T$. By increasing the
temperature,  at small deviations from the flat-band point, the entropy  approaches the
flat-band value  $s(J_2=1/2,T)$ (blue line in
figure~\ref{F8}).

\begin{figure}[!t]
\begin{center}
\includegraphics[width=.98\textwidth]{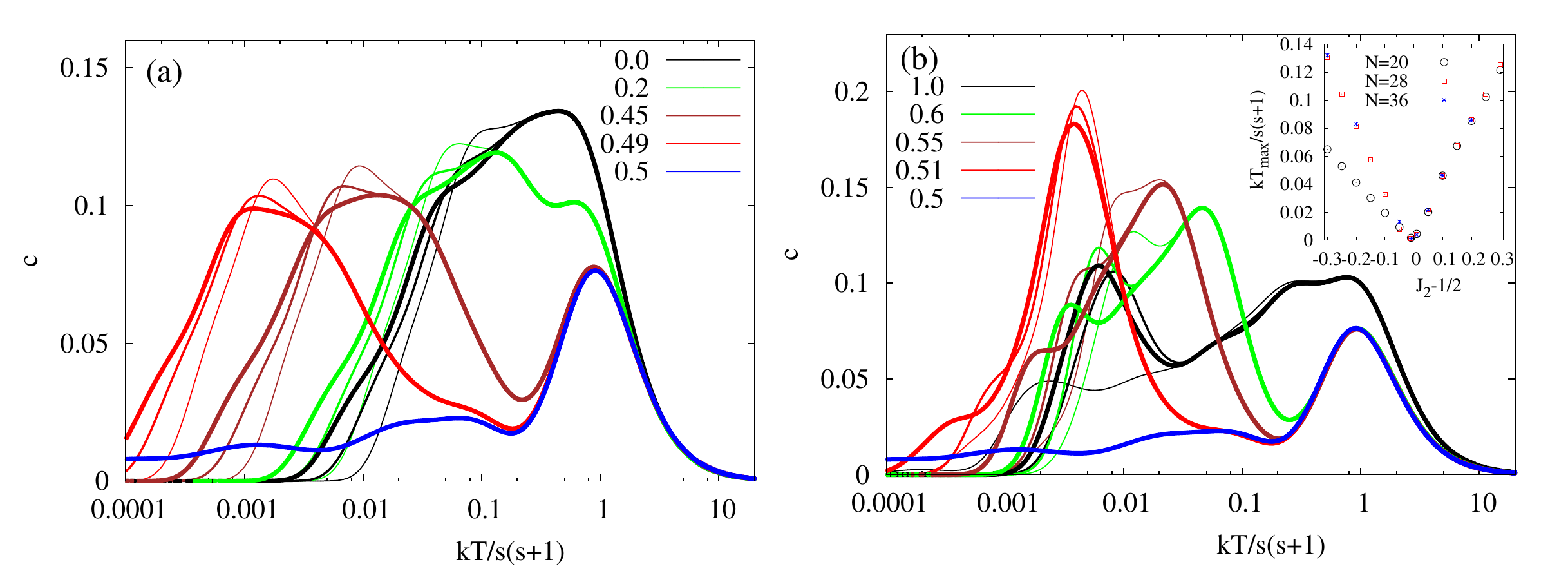}
\caption{(Colour online) Main panel: Specific heat per site 
of periodic chains of $N=20,28,36$ sites [$N=20$ (ED) --- thin,
$N=28$ (FTLM) --- medium, $N=36$ (FTLM) --- thick] 
for spin $s=1/2$, $J_1=-1$ and various values of (a) $J_2
\leqslant \kappa_\text{c}$ and (b) 
$J_2  \geqslant \kappa_\text{c}$ 
(the $J_2$ values are given in the legend)
at zero magnetic field.
The inset in panel (b) shows the position of the low-temperature maximum of
$c(T)$
as a function of the 
distance to the transition point $d_\text{f}=|J_2/J_1|-\kappa_\text{c}$.
}
\label{F9}
\end{center}
\end{figure}

The specific heat $c(T)$ is shown in figure~\ref{F9}~(a) and (b) for a few
values below and above the flat-band point $\kappa_\text{c}$. 
Apparently, $c(T)$ 
 exhibits clear signs of flat-band physics  in a sizeable parameter
region below and above  $\kappa_\text{c}$, namely a well-pronounced low-temperature peak  
coming from the former LMM ground states. On the other hand, the main peak
is quite stable against small deviations from $\kappa_\text{c}$. 
Noticeable  finite-size effects set in  
around the low-temperature peak, i.e., only at very low temperatures.
The  position of the low-temperature peak $T_{\rm max}$ depends on the energy-shift of the
LMM states, i.e., on the distance from the flat-band point
 $d_\text{f}$, see the inset in figure~\ref{F9}~(b).   
For $d_\text{f}>0$, there is a linear relation 
$T_{\rm max} \approx 0.33 d_\text{f}$, $0 < d_\text{f} \lesssim 0.3$, where the finite-size
effects are small.
On the other hand, for $d_\text{f}<0$, the peak position is noticeably dependent on
$N$, but there is no doubt of the double-maximum structure in the $c(T)$
profile.

\begin{figure}[!t]
\begin{center}
\includegraphics[width=.55\textwidth]{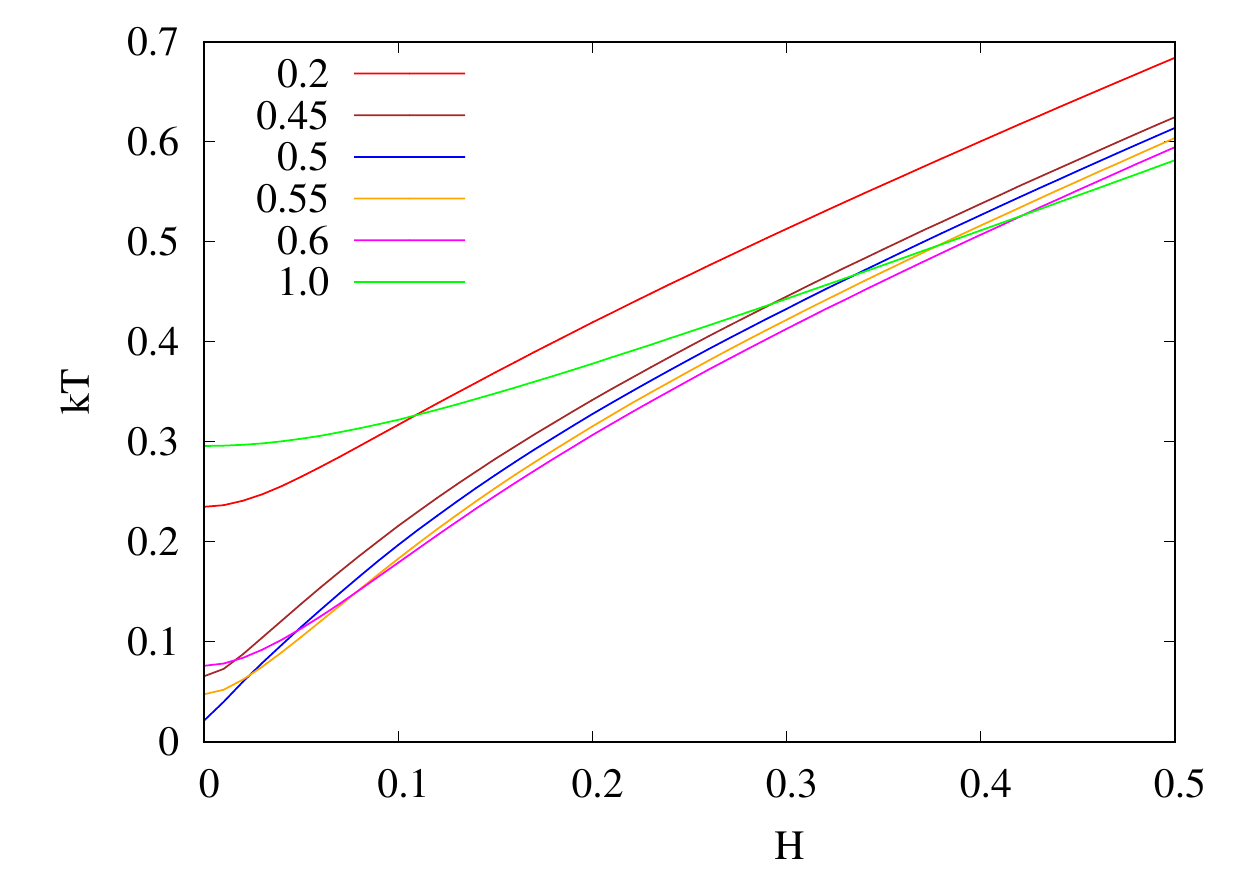}
\caption{(Colour online) Temperature in dependence 
on the
applied magnetic field for an isentropic cooling with fixed entropy $S/N=0.5$  
for spin $s=1/2$ and $N=36$ and various values of $J_2$ (given in the legend). 
Corresponding data for $N=20$ (not shown) demonstrate that finite-size effects
are small. 
}
\label{F10}
\end{center}
\end{figure}

Last but not least, we briefly discuss the magnetic cooling.
There are several theoretical studies reporting an enhanced magnetocaloric
effect in the vicinity of a quantum phase transition, in particular, if there
is a residual entropy at the transition point, see, e.g.,
references~\cite{6,9,12,63,64,65}.
However, most of the previous studies
in flat-band systems report  on an enhanced  magnetocaloric effect near the
saturation field \cite{6,9,12,63,64,65},
which often is not accessible in experiments.
By contrast, the  FM-AFM sawtooth chain exhibits this phenomenon when
approaching zero field \cite{15} which is more promising while thinking
in terms of a possible experimental realization.
In figure~\ref{F10} we show as an example the temperature variation as a function of the
applied magnetic field for an isentropic cooling with fixed entropy $S/N=0.5$
for spin $s=1/2$ and $N=36$ and various values of $J_2$.   
Apparently, there is a noticeable downturn in the $T(H)$ curve as $H \to 0$
for values of $J_2$ in the vicinity of the flat-band point, indicating the
presence of an
enhanced magnetocaloric effect in the FM-AFM sawtooth chain.

\section{Summary}
\label{sec6}
\setcounter{equation}{0}
\looseness=-1 In the present paper we study finite spin-$s$ Heisenberg 
sawtooth chains with   
ferromagnetic (FM) zigzag bonds $J_1<0$ and competing antiferromagnetic
(AFM) basal bonds
$J_2>0$ by means of full exact diagonalization and the finite-temperature
Lanczos
method.   
The model exhibits, at  $\kappa_\text{c}= |J_2/J_1|=1/2$,
a zero-temperature transition 
between a ferro- and a ferrimagnetic ground state.
At the transition 
(flat-band) point $\kappa_\text{c}$,
the lowest one-magnon excitation band from the ferromagnetic state is
flat (dispersionless) 
and  has zero energy. This  leads 
to a massively degenerate ground-state manifold 
resulting in a
residual entropy $\lim_{N \to \infty}  S_0(N)/N = \frac{1}{2}\ln 2$, that is
independent of $s$.
Moreover, in the extreme quantum case $s=1/2$ already for the finite systems
of up to $N=36$ sites, the excitations above  
the ground-state manifold are 
virtually gapless with the result that the finite-size data 
practically  correspond to the thermodynamic limit.

For spin quantum numbers $s>1/2$, the quantum effects at low temperatures
remain strong, even for the largest spin $s=3$ considered here. Thus, at the
flat-band point, the
specific-heat 
profile exhibits a well-pronounced maximum with a shoulder-like part and
long tail down to very low
temperatures below
this maximum for all $s=1,\ldots,3$, whereas this feature  is not present in the classical
case.  

In a sizeable parameter region around the flat-band point,
the former massively degenerate ground-state manifold acts as 
a large manifold of low-lying excitations setting an extra low-energy scale
yielding unconventional low-temperature thermodynamics which can be understood
as a remnant of flat-band physics.     

A specific feature of the flat-band system at hand is the strong influence of
an applied magnetic field on the low-temperature properties caused by a
partial lifting  of degenerate ground-state manifold, i.e., most states
collected in the residual entropy at zero field become low-energy
excitations according to their Zeeman energy.     

Finally, we argue that our results for the FM-AFM Heisenberg sawtooth chain
might be (at least to some extent) representative for other systems with a large 
residual
entropy such as the three-coloring $XXZ$ sawtooth chain \cite{17,18,23} and
the FM-AFM kagome chain \cite{66}.

\section*{Acknowledgements}
JR is deeply indebted to Oleg Derzhko for many
inspiring scientific discussions and a very fruitful collaboration
over more than 20 years.
All authors thank Oleg Derzhko for useful discussions and Andreas
Honecker for critical comments.
J.R. and J.S. thank the DFG for financial support (grants RI 615/25-1 and
SCHN 615/28-1).

\newpage
\ukrainianpart

\title{Аномальна термодинаміка квантової спінової системи з великою залишковою ентропією}
\author{Й. Ріхтер \refaddr{label1,label2}, Й. Шуленбурґ \refaddr{label3}, Д. В. Дмітрієв \refaddr{label4}, В. Я. Крівнов \refaddr{label4}, Ю. Шнак  \refaddr{label5}}
\addresses{
\addr{label1} Інститут фізики, Університет Магдебурга, поштова скринька 4120, Магдебург 39016, Німеччина
\addr{label2} Інститут Макса Планка фізики складних систем, Ньотнітцерштрасе 38, D--01187 Дрезден, Німеччина
\addr{label3} Університетський обчислювальний центр, Університет Магдебурга, поштова скринька 4120, Магдебург 39016, Німеччина
\addr{label4} Інститут біохімічної фізики РАН, вул. Косигіна 4, Москва 119334, Росія
\addr{label5} Університет Білєфельда, факультет фізики, поштова скринька 100131, D--33501 Білєфельд, Німеччина
}

\makeukrtitle

\begin{abstract}
На відміну від класичних сильно фрустрованих систем, їхні квантові відповідники зазвичай мають невироджений основний стан. Контрприкладом є знаменитий спіновий пилкоподібний ланцюжок Гайзенберґа з феромагнітними зигзаг зв'язками $J_1$ і конкуруючими антиферомагнітними базисними зв'язками $J_2$. У точці квантового фазового переходу $|J_2/J_1|=1/2$ така модель виявляє плоску зону одномагнонних збуджень, яка веде до сильно виродженого многовиду основних станів, що призводить до великої залишкової ентропії. Отже, для спін-1/2 моделі залишкова ентропія становить точно половину від її максимального значення $\lim_{T\to\infty} S(T)/N = \ln2$. В даній роботі ми детально вивчаємо роль спінового квантового числа $s$ і магнітного поля $H$ в області параметрів біля точки переходу (плоска зона). Для цього ми використовуємо повну діагоналізацію до $N=20$ вузлів ґратки, а також скінченно-температурний метод Ланцоша до $N=36$ вузлів, щоб обчислити густину станів, а також температурну залежність теплоємності, ентропії та сприйнятливості. Дослідження ланцюжків довжиною до $N=36$ дозволяє акуратний скінченно-розмірний аналіз. В плоскозонній точці ми знайшли надзвичайно малий скінченно-розмірний ефект для спіна $s=1/2$, тобто числові дані віртуально відповідають термодинамічній границі. В усіх інших випадках скінченно-розмірні ефекти є все ще малі і стають помітними при дуже низьких температурах. У значній області параметрів навколо плоскозонної точки вплив попереднього сильно виродженого многовиду основних станів діє як великий многовид низькоенергетичних збуджень, який в точці переходу, а також в її околі, веде до незвичайних термодинамічних властивостей, таких як додатковий низькотемпературний максимум у теплоємності. До того ж, існує дуже сильний вплив магнітного поля на низькотемпературну термодинаміку, включаючи посилений магнетокалоричний ефект.
\keywords квантова модель Гайзенберґа, фрустрація, пилкоподібний ланцюжок, залишкова ентропія

\end{abstract}


\begin{thebibliography}{99}
\bibitem{1} Sen  D., Shastry B.S., Walstedt R.E., Cava R., Phys. Rev. B, 1996, {\bf 53}, 6401, \doi{10.1103/PhysRevB.53.6401}.  
                                
\bibitem{2} Schulenburg J., Honecker A., Schnack J., Richter J., Schmidt H.-J., 
      Phys. Rev. Lett., 2002,  {\bf 88}, 167207, \doi{10.1103/PhysRevLett.88.167207}.      
       
\bibitem{3} Chen S., B\"uttner H., Voit J., 
Phys. Rev. B, 2003,  {\bf 67}, 054412, \doi{10.1103/PhysRevB.67.054412}. 
                               
\bibitem{4}  Blundell S.A., N\'u\~nez-Regueiro M.D., 
Eur. Phys. J. B, 2003,  {\bf  31}, 453, \doi{10.1140/epjb/e2003-00054-2}.  
 
\bibitem{5} Blundell S.A.,  N\'u\~nez-Regueiro M.D.,  
     J. Phys.: Condens. Matter, 2004, {\bf 16}, S791,\\ \doi{10.1088/0953-8984/16/11/031}. 
           
\bibitem{6} Zhitomirsky M.E., Honecker A., 
    J. Stat. Mech.: Theory Exp., 2004, {\bf 2004}, P07012,\\ \doi{10.1088/1742-5468/2004/07/P07012}.   
                         
\bibitem{7} Tonegawa T., Kaburagi M., 
J. Magn. Magn. Mater., 2004,  {\bf 272--276},  898, \doi{10.1016/j.jmmm.2003.11.367}.  
          
\bibitem{8} Chandra V.R., Sen D., Ivanov N.B., Richter J., 
Phys. Rev. B, 2004, {\bf 69}, 214406,\\ \doi{10.1103/PhysRevB.69.214406}.  
           
\bibitem{9} Zhitomirsky M.E., Tsunetsugu H., 
Phys. Rev. B, 2004, {\bf  70}, 100403, \doi{10.1103/PhysRevB.70.100403}.

\bibitem{10} Derzhko O., Richter J., 
Phys. Rev. B, 2004, {\bf 70}, 104415, \doi{10.1103/PhysRevB.70.104415}.

\bibitem{11} Kaburagi M., Tonegawa T., Kang M., 
J. Appl. Phys., 2005, {\bf 97}, 10B306, \doi{10.1063/1.1851893}.

\bibitem{12} Derzhko O., Richter J., 
Eur. Phys. J. B, 2006, {\bf 52}, 23, \doi{10.1140/epjb/e2006-00273-y}.

\bibitem{13} Richter J., Derzhko O., Honecker A., 
     Int. J.  Mod. Phys. B, 2008, {\bf  22}, 4418, \doi{10.1142/S0217979208050176}.
     
\bibitem{14} Hao Z., Wan Y., Rousochatzakis I.,
     Wildeboer J., Seidel A., Mila F., Tchernyshyov O.,
Phys. Rev. B, 2011, {\bf 84}, 094452, \doi{10.1103/PhysRevB.84.094452}.

\bibitem{15} Krivnov V.Ya., Dmitriev D.V., Nishimoto S.,
     Drechsler S.-L., Richter J., 
Phys. Rev. B, 2014, {\bf  90}, 014441, \doi{10.1103/PhysRevB.90.014441}.

\bibitem{16} Dmitriev D.V., Krivnov V.Ya., 
Phys. Rev. B, 2015, {\bf  92}, 184422, \doi{10.1103/PhysRevB.92.184422}.

\bibitem{17} Changlani H.J., Kochkov D., Kumar K.,
     Clark B.K., Fradkin E., 
Phys. Rev. Lett., 2018, {\bf 120}, 117202, \doi{10.1103/PhysRevLett.120.117202}.

\bibitem{18} Changlani H.J., Pujari S., Chung C.-M.,
    Clark  B.K., Phys. Rev. B, 2019, {\bf 99}, 104433,\\ \doi{10.1103/PhysRevB.99.104433}.
    
\bibitem{19} Dmitriev D.V., Krivnov V.Ya., Richter J., Schnack J., 
Phys. Rev. B, 2019, {\bf  99}, 094410,\\ \doi{10.1103/PhysRevB.99.094410}.

\bibitem{20} Yamaguchi T., Drechsler S.-L.,
     Ohta Y., Nishimoto S., 
Phys. Rev. B, 2020, {\bf 101}, 104407,\\ \doi{10.1103/PhysRevB.101.104407}.

\bibitem{21} Metavitsiadis A., Psaroudaki C., Brenig W.,
Phys. Rev. B, 2020, {\bf 101},  235143,\\ \doi{10.1103/PhysRevB.101.235143}.

\bibitem{22} Dmitriev D.V., Krivnov V.Ya., Schnack J., 
     Richter J., 
Phys. Rev. B, 2020, {\bf 101}, 054427,\\ \doi{10.1103/PhysRevB.101.054427}.

\bibitem{23} Derzhko O., Schnack J., Dmitriev D.V., Krivnov V.Ya.,
     Richter J., Eur. Phys. J. B, 2020, \textbf{93}, 161, \doi{10.1140/epjb/e2020-10224-1}.
     
\bibitem{24} McClarty P.A., Haque M., Sen A., Richter J.,
     Phys. Rev. B, 2020,  {\bf 102}, 224303,
    \\\doi{10.1103/PhysRevB.102.224303}.
     
\bibitem{25} Tasaki H., 
Phys. Rev. Lett., 1992, {\bf 69}, 1608, \doi{10.1103/PhysRevLett.69.1608}.

\bibitem{26} Mielke A., Tasaki H., 
Commun. Math. Phys., 1993, {\bf 158}, 341, \doi{10.1007/BF02108079}.

\bibitem{27} Tasaki H., 
      J. Stat. Phys., 1996, {\bf  84}, 535, \doi{10.1007/BF02179652}.
      
\bibitem{28} Watanabe Y.,  Miyashita S., 
J. Phys. Soc. Jpn., 1997, {\bf  66}, 2123, \doi{10.1143/JPSJ.66.2123}.

\bibitem{29} Arita R., Shimoi Y., Kuroki K.,  Aoki H., 
Phys. Rev. B, 1998, {\bf 57}, 10609, \doi{10.1103/PhysRevB.57.10609}.

\bibitem{30} Derzhko O., Richter J., Honecker A., Maksymenko M.,
     Moessner R., 
Phys. Rev. B, 2010, {\bf  81}, 014421, \doi{10.1103/PhysRevB.81.014421}.

\bibitem{31} Maksymenko M., Honecker A., Moessner R., Richter J.,  Derzhko O., 
Phys. Rev. Lett., 2012,  {\bf 109}, 096404, \doi{10.1103/PhysRevLett.109.096404}.

\bibitem{32} Flach S., Leykam D., Bodyfelt J.D.,
     Matthies P.,  Desyatnikov A.S., 
EPL, 2014, {\bf  105}, 30001,\\ \doi{10.1209/0295-5075/105/30001}.

\bibitem{33} Liu R., Nie W., Zhang W., Sci. Bull., 2019, {\bf 64}, 1490, \doi{10.1016/j.scib.2019.08.013}.  
                      
\bibitem{34} Weimann S., Morales-Inostroza L., Real B.,   
     Cantillano C., Szameit A.,  Vicencio R.A., 
Opt. Lett., 2016, {\bf  41}, 2414, \doi{10.1364/OL.41.002414}.    
                          
\bibitem{35} Tang L., Song D., Xia S., Xia S.,  
     Ma J., Yan W., Hu Y., Xu J., Leykam D.,  Chen Z., 
Nanophotonics, 2020,  {\bf  9}, 1161, \doi{10.1515/nanoph-2020-0043}.

\bibitem{36} Ruiz-P\'erez C., Hern\'andez-Molina M.,      
     Lorenzo-Luis P., Lloret F., Cano J.,  Julve M., 
Inorg. Chem., 2000,  {\bf  39}, 3845, \doi{10.1021/ic000314n}.    
                          
\bibitem{37} Inagaki Y., Narumi Y., Kindo K.,     
     Kikuchi H., Kamikawa T., Kunimoto T.,            
     Okubo S., Ohta H., Saito T.,               
      Azuma M., Takano M., Nojiri H.,             
     Kaburagi M.,  Tonegawa T., 
J. Phys. Soc. Jpn., 2005,  {\bf  74}, 2831, \doi{10.1143/JPSJ.74.2831}.
          
\bibitem{38} Baniodeh A., Magnani N., Lan Y., Buth G., Anson C.E.,   
     Richter J., Affronte M., Schnack J., Powell A.K.,   
npj Quantum Mater., 2018,  {\bf  3}, 10, \doi{10.1038/s41535-018-0082-7}.      
                            
\bibitem{39} Heinze L., Jeschke H., Metavitsiadis A., Reehuis M.,     
     Feyerherm R., Hoffmann J.-U., Wolter A.U.B., Ding~X.,  
     Zapf V., Moya C.C., Weickert F., Jaime M.,        
     Rule K.C., Menzel D., Valent\'i R., Brenig W.,  S\"ullow S., 
      Preprint~\arxiv{1904.07820}, 2019.  
                          
\bibitem{40} Nhalil H., Baral R., Khamala B.O., Cosio A.,      
     Singamaneni S.R., Fitta M., Antonio D., Gofryk K.,      
     Zope R.R., Baruah T., Saparov B., Nair H.S., 
 Phys. Rev. B, 2019,   {\bf 99}, 184434,
 \doi{10.1103/PhysRevB.99.184434}.    
          
\bibitem{41} Derzhko O., Richter J., Maksymenko M., 
Int. J. Mod. Phys. B, 2015,  {\bf 29}, 1530007,\\ \doi{10.1142/S0217979215300078}. 
                                                 
\bibitem{42} Kikuchi H., Fujii Y., Chiba M., Mitsudo S., Idehara T., Tonegawa T.,
 Okamoto K., Sakai T., Kuwai T.,  Ohta H., 
Phys. Rev. Lett., 2005,  {\bf 94}, 227201, \doi{10.1103/PhysRevLett.94.227201}.  
                                                
\bibitem{43} Honecker A., Hu S., Peters R., Richter J., 
     J. Phys.: Condens. Matter, 2011,  {\bf 23}, 164211,\\ \doi{10.1088/0953-8984/23/16/164211}.
       
\bibitem{44} Okuma R., Nakamura D., Okubo T., Miyake A.,              
Matsuo A., Kindo K., Tokunaga M., Kawashima  N.,                                                               
     Takeyama~S., Hiroi Z., 
Nat. Commun., 2019,  {\bf 10}, 1229, \doi{10.1038/s41467-019-09063-7}.

\bibitem{45} Tanaka H., Kurita N., Okada M.,                                                                   
      Kunihiro E., Shirata Y., Fujii K., Uekusa H., Matsuo A., Kindo K., Nojiri H., 
J. Phys. Soc. Jpn., 2014,  {\bf  83}, 103701, \doi{10.7566/JPSJ.83.103701}. 
                               
\bibitem{46} Richter J., Krupnitska O., Baliha V., Krokhmalskii T.,  Derzhko O., 
Phys. Rev. B, 2018,  {\bf 97}, 024405, \doi{10.1103/PhysRevB.97.024405}.
        
\bibitem{47} Shirakami R., Ueda H., Jeschke H.O.,         
      Nakano H., Kobayashi S., Matsuo A.,        
      Sakai T., Katayama N., Sawa H.,      
      Kindo K., Michioka C., Yoshimura K.,
Phys. Rev. B, 2019,  {\bf  100}, 174401, \doi{10.1103/PhysRevB.100.174401}.

\bibitem{48} L\"auchli A.M., In: Introduction to Frustrated Magnetism,
      Mila F., Lacroix C., Mendels~P.~(Eds.), Springer,
      Berlin, Heidelberg, 2011, 34.
      
\bibitem{49} Schulenburg J., Spinpack-2.59, Magdeburg University, 2019.

\bibitem{50} Richter J.,  Schulenburg J., 
Eur. Phys. J. B, 2010,  {\bf  73}, 117, \doi{10.1140/epjb/e2009-00400-4}.

\bibitem{51} Jakli\v{c} J.,  Prelov\v{s}ek P., 
Phys. Rev. B, 1994, {\bf   49}, 5065, \doi{10.1103/PhysRevB.49.5065}.

\bibitem{52} Jakli\v{c} J.,  Prelov\v{s}ek P., 
Adv. Phys., 2000,  {\bf 49}, 1, \doi{10.1080/000187300243381}.

\bibitem{53} Hams A., De Raedt H., 
Phys. Rev. E, 2000,  {\bf  62}, 4365, \doi{10.1103/PhysRevE.62.4365}.

\bibitem{54} Schnack J.,  Wendland O., 
Eur. Phys. J. B, 2010,  {\bf 78}, 535, \doi{10.1140/epjb/e2010-10713-8}.

\bibitem{55} Prelov\v{s}ek P.,  Bon\v{c}a J., In: Strongly Correlated Systems:
      Numerical Methods,  Avella A., Mancini~F.~(Eds.), Springer, Berlin, Heidelberg, 2013, 1--30.  
            
\bibitem{56} Hanebaum O.,  Schnack J., 
Eur. Phys. J. B, 2014,  {\bf   87}, 194, \doi{10.1140/epjb/e2014-50360-5}.

\bibitem{57} Schmidt B., Thalmeier P., 
Phys. Rep., 2017,  {\bf  703}, 1, \doi{10.1016/j.physrep.2017.06.004}.

\bibitem{58}  Prelov\v{s}ek P., In: The Physics of Correlated Insulators, Metals, and Superconductors, Pavarini E., Koch E., Scalettar R., Martin~R.M.~(Eds.), Forschungszentrum J\"ulich GmbH, J\"ulich, 2017, 7.1--7.23.

\bibitem{59} Schnack J., Schulenburg J., Richter J., 
Phys. Rev. B, 2018,  {\bf  98}, 094423, \doi{10.1103/PhysRevB.98.094423}.

\bibitem{60} Schnack J., Richter J.,  Steinigeweg R., 
      Phys. Rev. Res., 2020,  {\bf   2}, 013186,\\ \doi{10.1103/PhysRevResearch.2.013186}.
      
\bibitem{61} Seki K., Yunoki S., 
Phys. Rev. B, 2020,  {\bf  101}, 235115, \doi{10.1103/PhysRevB.101.235115}.

\bibitem{61a} Schnack J., Schulenburg J., Honecker A.,  Richter J.,
Phys. Rev. Lett., 2020, \textbf{125}, 117207,\\ \doi{10.1103/PhysRevLett.125.117207}.

\bibitem{62} Lohmann A., Schmidt H.-J.,  Richter J., 
Phys. Rev. B, 2014,  {\bf   89}, 014415, \doi{10.1103/PhysRevB.89.014415}.

\bibitem{63} Zhitomirsky M.E., 
Phys. Rev. B, 2003,  {\bf  67}, 104421, \doi{10.1103/PhysRevB.67.104421}.

\bibitem{64} Schnack J., Schmidt R.,  Richter J.,
 Phys. Rev. B, 2007,  {\bf   76}, 054413, \doi{10.1103/PhysRevB.76.054413}.
 
\bibitem{65} Wolf B., Honecker A., Hofstetter W.,
        Tutsch U.,  Lang M., 
        Int. J.  Mod. Phys. B, 2014,  {\bf  28}, 1430017, \doi{10.1142/S0217979214300175}.
        
\bibitem{66}  Dmitriev D.V., Krivnov V.Ya.,
J. Phys.: Condens. Matter, 2017,  {\bf  29}, 215801, \doi{10.1088/1361-648X/aa68f6}.


\end{thebibliography}
\end{document}